\documentclass[aps,prb,preprint,unsortedaddress,showpacs]{revtex4}
\usepackage{amssymb}
\usepackage{graphicx}

\begin{document}

\title{Carbon Based Superconductors}

\author{Reinhard K. Kremer$^*$}
\author{Jun Sung Kim}
\author{Arndt Simon}
\affiliation{Max-Planck-Institut f\"ur Festk\"orperforschung,
Heisenbergstr.1, D-70569 Stuttgart, Germany}

\date{\today}

\vspace*{2cm}

 \begin{abstract}
We review the characteristics of some carbon based novel
superconductors which emerged in the past two decades since the
discovery of superconductivity in the high-$T_{\rm c}$ oxocuprates.
In particular, we summarize the properties of ternary layered
carbide halides of the rare rarth metals with composition
RE$_2$C$_2$X$_2$ (RE = Y, La; X=Cl, Br, I) and of the rare earth di-
and sesquicarbides, YC$_2$, LaC$_2$ and La$_2$C$_3$. Finally, we
briefly discuss the properties of the recently discovered Ca and Yb
intercalated graphite superconductors, CaC$_6$ and YbC$_6$.

\end{abstract}

\pacs{abc}

\maketitle

{\section{Introduction}}

The discovery of high-$T_{\rm c}$ superconductivity by Bednorz and
M\"uller\cite{Bednorz} in 1986  marks the beginning of a period of
a vivid search for - chemically and physically partly extremely
complex - new oxocuprates and for theoretical approaches to
understand their puzzling properties, quite a few of which
remained controversial even until today. The advent of this
completely unexpected class of new superconductors also revived
the interest in more conventional - `low-$T_{\rm c}$' -
superconductors. In due course, a number of new  systems were
found, and already known superconductors were reinvestigated with
improved and refined experimental and theoretical tools. These
activities led to surprising new discoveries as that of the 40 K
superconductor MgB$_2$ by Nagamatsu \textit{et
al.}\cite{Nagamatsu} Apart from its $T_{\rm c}$, MgB$_2$ is
special primarily for two reasons: Compared \textit{e.g.} to the
high-$T_{\rm c}$ oxocuprates its crystal structure is of
remarkable simplicity allowing electronic and phononic structure
calculations of high precision, and MgB$_2$ is the first system
for which \textit{multigap} superconductivity has independently
been evidenced by several experimental
techniques.\cite{Bouquet,Crabtree,Dolgov}

Until the discovery of MgB$_2$, doped fullerenes had shown the
highest $T_{\rm c}$ values after the high-$T_{\rm c}$ oxocuprates.
 With large enough quantities of purified C$_{60}$
available, \cite{Kratschmer} Hebard \textit{et al.} prepared
superconductors with a $T_{\rm c}$ of 18 K by doping
polycrystalline C$_{60}$ and C$_{60}$ films with alkali
metals.\cite{Hebard} Subsequently, by adjusting the separation of
the C$_{60}$ molecules using a proper composition of different
alkali metals, $T_{\rm c}$'s up to $\sim$33 K were
reached.\cite{Tanigaki}

Superconductivity in doped fullerenes also redraw attention to
carbon based superconductors in general. Especially,  binary and
quasibinary transition metal carbides have a long history in showing
$T_{\rm c}$'s which were among the highest found before the
discovery of the high-$T_{\rm c}$ oxocuprates. \cite{Roberts} Later
borocarbides of composition REM$_2$B$_2$C, with RE = Y or Lu and M =
Ni or Pd, with $T_{\rm c}$'s up to 22~K attracted considerable
interest. \cite{Nagarajan,Cava}

Superconductivity in graphite intercalation compounds (GICs) is
another early field of research which recently was revived. The
discovery of superconducting GICs dates back to the pioneering
work of Bernd Matthias' group in the 1960's, however, the $T_{\rm
c}$'s of these early GICs remained well below 1~K.
\cite{Hannay,Dresselhaus} Subsequently, the $T_{\rm c}$'s of
alkali metal intercalated GICs could be raised by intercalation
under pressure with \textit{e.g.} Li and Na, but $T_{\rm c}$ did
not significantly exceed the boiling point of liquid helium.
\cite{Belash} It was not until recently that $T_{\rm c}$ of the
GICs could be significantly enhanced by intercalating divalent
alkaline earth metals like Ca and Yb. \cite{Weller}

Finally, after graphite and C$_{60}$, diamond was also converted
into a superconductor by hole doping induced by a substitution of
about 3\%  B into C sites. Ekimov \textit{et al.} showed that such
a boron-doped diamond is a bulk, type-II superconductor below
$T_c$\,$\sim$\,4 K with superconductivity surviving in a magnetic
field up to $H_{\rm{c2}}$(0)\,$\geq$\,3.5 T. \cite{Ekimov}

In our search for complex metal-rich rare earth halides we found a
series of new superconducting layered carbide halides of the rare
earth metals with $T_{\rm c}$'s up to $\sim$10~K.
\cite{Simon1991,ZAAC,HennPRL} For a deeper understanding of the
chemistry and physics of these we in turn reinvestigated also the
properties of binary dicarbides and sesquicarbides of composition
REC$_2$ and RE$_2$C$_3$, with R\,=\,Y,La. Superconductivity in
binary  carbides of rare earth metals had been an intensively
investigated topic in the sixties and seventies of the last
century. In this family of compounds $T_{\rm c}$ values peaked
with (Y$_{0.7}$Th$_{0.3})_2$C$_3$ at 17 K.\cite{Krupka69}

Superconductivity in rare earth metal sesquicarbides recently
regained considerable attention after the reports by Amano
\textit{et al.} and Nakane \textit{et al.} about the successful
synthesis of binary Y$_2$C$_3$ under high pressure conditions
($\sim$5~GPa).\cite{Amano,Nakane} The reported  $T_{\rm c}$'s
reached 18~K and the upper critical field exceeded 30~T.  In the
following we will summarize some of the characteristic properties
of the ternary layered rare earth metal carbide halides and the
binary di- and sesquicarbides. We conclude with some remarks on
our results
on the recently discovered alkali earth GICs.\\

\section{Superconductivity in Rare Earth Carbide Halides and Rare Earth Carbides}

\subsection{Ternary Layered Cabide Halides of the Rare Earth Metals}

\begin{figure}[h]
\includegraphics[width=8cm,angle=270]{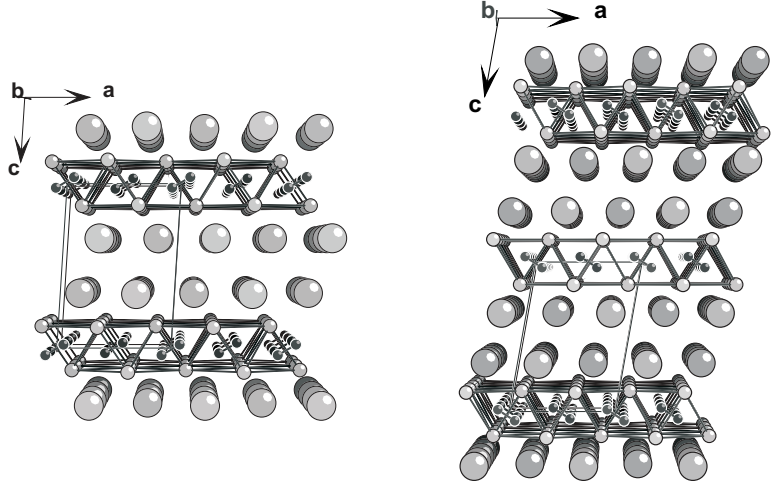}
\caption{(left) Crystal
structure of Y$_2$C$_2$I$_2$ (1s stacking
variant) and (right) crystal structure of  Y$_2$C$_2$Br$_2$ (3s
stacking variant) projected along [010] with the unit cells
outlined. C, Y, and (I,Br) atoms are displayed with increasing
size.} \label{Y2STR}
\end{figure}

The carbide halides of the rare earth metals, RE$_2$C$_2$X$_2$
(X\,=\,Cl, Br, I and RE being a rare-earth metal) crystallize with
layered structures which contain double layers of close-packed
metal atoms  sandwiched by layers of halogen atoms to form
X-RE-C$_2$-RE-X slabs as elementary building blocks. These connect
via van der Waals forces in stacks along the crystallographic
$c$-axis. Different stacking sequences (1s and 3s stacking
variants) have been found. The carbon atoms form C-C dumbbells
which occupy the octahedral voids  in the close-packed metal atom
doublelayers  (cf. Fig. \ref{Y2STR}).\cite{Schuller85,Mattausch92}

\begin{figure}[hpb]
\includegraphics[width=8cm,angle=270]{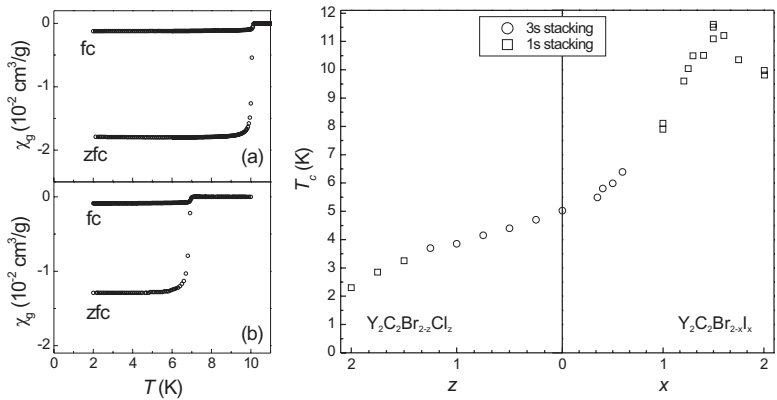} \caption{(left) field-cooled (fc)
and zero field-cooled (zfc) magnetic susceptibilities of (a)
Y$_2$C$_2$I$_2$ (after ref.[\onlinecite{Kremer2004}]) and (b)
La$_2$C$_2$Br$_2$ (after ref.[\onlinecite{Ahn99}]). (right) $T_{\rm
c}$'s of a series of quasiternary mixtures of
Y$_2$C$_2$Br$_{2-{\rm{x}}}$I$_{\rm{x}}$ and
Y$_2$C$_2$Br$_{2-{\rm{z}}}$Cl$_{\rm{z}}$. Different stacking
variants of the compounds are indicated by different symbols (after
ref.[\onlinecite{ZAAC}]).} \label{Y2CHI}
\end{figure}

Compounds containing the nonmagnetic rare-earth metals Y and La
are superconductors (Fig. \ref{Y2CHI}). The maximum $T_{\rm c}$ of
11.6 K which was achieved by adjusting the composition in the
quasi-ternary phases Y$_2$C$_2$(X,X')$_2$. \cite{ZAAC} The
variation of $T_c(x)$ across the transition of the 3s and the 1s
stacking variant indicating that superconductivity is essentially
a property of the configuration of an individual X-RE-C$_2$-RE-X
slab rather than of the stacking details in the crystal structure.
The transition temperatures of all known superconducting phases
RE$_2$C$_2$X$_2$ are compiled in Table \ref{TableTc}.

\begin{table}[ht]
\begin{tabular}{| c  |  c  |  c  | c |}
\hline
compound & \,\,  $T_{\rm c}$ (K)  \,\, & \,\,$\mu_{\rm{0}} H_{\rm{c2}}$ (T) \,\,& \,\,reference\,\,\\
\hline
$\rm Y_2C_2Cl_2$  & 2.3 & - &\onlinecite{ZAAC} \\
$\rm Y_2C_2Br_2$  & 5.04 & 3& \onlinecite{ZAAC,HennPRL,HennThesis} \\
$\rm Y_2C_2I_2$  & 10.04 & 12&\onlinecite{ZAAC,AhnDavies,Kremer2004,HennThesis} \\
$\rm Y_2C_2Br_{0.5}I_{1.5}$  & 11.6 & - &\onlinecite{ZAAC} \\
$\rm La_2C_2Br_2$  & 7.03 & - &\onlinecite{Ahn99} \\
$\rm La_2C_2I_2$   &1.72  & - &\onlinecite{Ahn99} \\
\hline
\end{tabular}
\caption{Transition temperatures and upper critical fields,
$\mu_{\rm{0}} H_{\rm{c2}}$,  of the known superconducting phases
RE$_2$C$_2$X$_2$ (RE\,=\,Y, La; X=Cl, Br, I)} \label{TableTc}
\end{table}

The heat capacity of Y$_2$C$_2$I$_2$ shows a sharp anomaly,
however with a jump height $\Delta C_P(T_{\rm C})$/$\gamma\,T_{\rm
C}$\,$\approx $\,2 which is considerably larger than the value
1.43 expected from weak coupling BCS theory.
\cite{HennPRL,Kremer2004}  A fit of the heat capacity anomaly with
the empirical $\alpha$-model \cite{Padamsee} indicates strong
coupling with 2$\Delta$(0)/k$_{\rm B}\,T_{\rm C} \approx $4.2, the
superconducting gap being enhanced by about 20\,\% over the BCS
value, similar to the $\sigma$\,-\,gap in MgB$_2$ \cite{Golubov}.
There is, however,  no indication from the temperature dependence
of the heat capacity anomaly for a multiple gap scenario. Using
approximate equations for strong coupling superconductors which
relate 2$\Delta$(0)/k$_{\rm B}\,T_{\rm C}$ and $\Delta C_P(T_{\rm
C}$)/$\gamma\,T_{\rm C}$ to the logarithmic average over the
phonon frequencies  $\omega_{ln}$ \cite{Carbotte,Allen} one
estimates the typical phonon frequency range for Y$_2$C$_2$I$_2$
to be $\sim$80\,-\,100\,cm$^{-1}$. In this range $A_g$ modes have
been discerned by Raman spectroscopy in which Y and halogen atoms
vibrate in-phase parallel and perpendicular to the layers.
\cite{Puschnig,HennRaman}

C stretching and tilting vibrations have considerably higher
energies, and their role for electron-phonon coupling,
particularly in the case of the tilting modes, could be important.
\cite{Simon97} The electronic structure in close neighborhood to
the Fermi energy, $E_{\rm{F}}$, is characterized by bands of low
dispersion which are reminiscent of the quasimolecular character
of the HOMO and LUMO orbitals of an isolated C-C dumbbell.
\cite{Puschnig,Miller} These together with highly dispersive bands
establish a flat/steep band scenario which in our view is a
prerequisite of superconductivity in a more general sense.
\cite{Deng}

The low-dispersive bands give rise to two peaks in the electronic
density of states, DOS, each about 100 meV above and below the
Fermi energy which enclose a `pseudogap' at $E_{\rm{F}}$.
\cite{Puschnig,Miller} Deviations from the linear temperature
dependence of the Korringa relaxation of  $^{13}$C nuclei probed
by $^{13}$C NMR are a clear manifestation for the proposed
structure in the DOS close to $E_{\rm{F}}$. \cite{Herrling}

The electronic structure and the dispersion of the bands in the
vicinity of $E_{\rm{F}}$  is very sensitive to slight structural
variations and can be very effectively tuned e.g. by hydrostatic
pressure to increase the DOS and maximize $T_{\rm c}$.
\cite{AhnPress2} When hydrostatic pressure is applied to
Y$_2$C$_2$I$_2$ $T_{\rm c}$ increases,  and a maximum of about
11.7 K is reached at 2 GPa, similar to the maximum $T_{\rm c}$
found in the quasi-ternary mixtures.
\cite{HennPress,AhnDavies,AhnPress2} The increase of $T_{\rm c}$
with pressure Y$_2$C$_2$I$_2$ and also La$_2$C$_2$Br$_2$ is
remarkable and parallels the findings in observed for the Hg based
oxocuprates but also for $fcc$-La for which similar values for the
relative increase 1/$T_{\rm c}$\,$\cdot$\,$dT_{\rm c}$\,/\,$dP$,
have been detected. \cite{Gao,Schilling2}

\begin{figure}[h]
\includegraphics[width=8cm,angle=270]{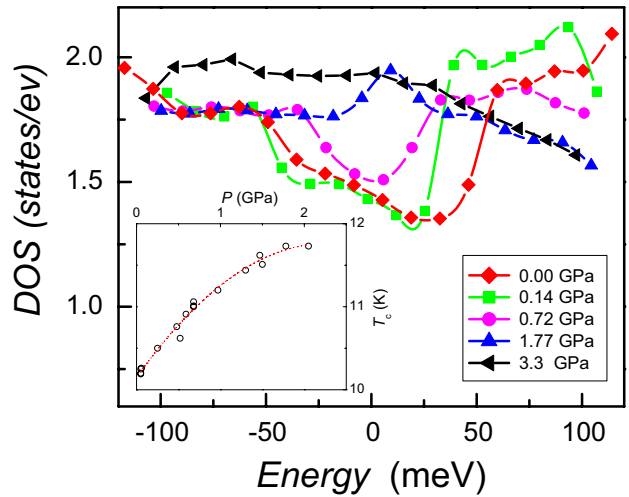} \caption{Electronic density of
states, DOS, of Y$_2$C$_2$I$_2$ in the close vicinity to
$E_{\rm{F}}$. The inset shows the pressure dependence of $T_{\rm c}$
of Y$_2$C$_2$I$_2$. The dotted line is a guide to the eye (after
ref. [\onlinecite{AhnPress2,HennPress}]).} \label{Y2PRES}
\end{figure}

\subsection{Binary Dicarbides and Sesquicarbides of the Rare Earth Metals}

\begin{figure}[h]
\includegraphics[width=8cm,angle=270]{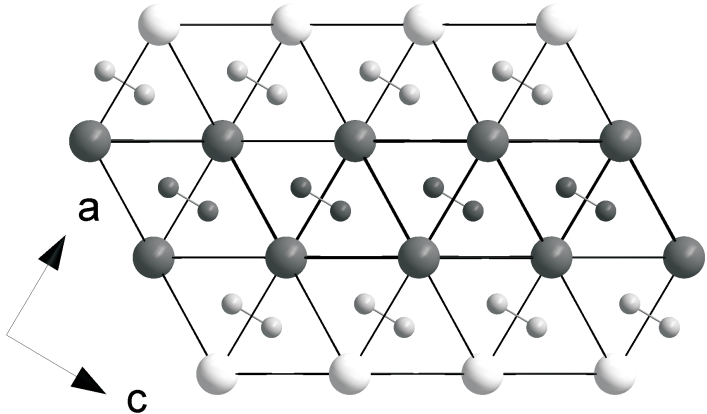}

\caption[]{Crystal structure of YC$_2$ along [0 1 0]. Y and C atoms
are drawn with decreasing size. An Y\,-\,C$_2$\,-\,Y doublelayer as
found in the ternary carbide halides of the rare earths metals,
RE$_2$C$_2$X$_2$ (RE\,=\,Y, La; X\,=\,Cl, Br, I), is highlighted in
dark grey.} \label{YC2STR}
\end{figure}

YC$_2$ crystallizes with the body centered tetragonal CaC$_2$
structure type (Fig. \ref{YC2STR}) with C-C dumbbells centering Y
metal atom octahedra which are slightly elongated along
[001].\cite{Atoji} YC$_2$ had been found to be a superconductor
with a $T_{\rm c}$\,$\sim$\,3.88 K. \cite{Giorgi} Proper heat
treatment of stoichiometric $\rm YC_2$ samples results in
superconductors with a sharp transition and onset $T_{\rm c}$'s up
to 4.02(5)\,K , somewhat increased over those previously reported.
\cite{GuldenPRB,GuldenThesis} LaC$_2$ shows a $T_{\rm c}$ of about
1.6 K. \cite{Giorgi}

\begin{figure}[h]
\includegraphics[width=8cm,angle=0]{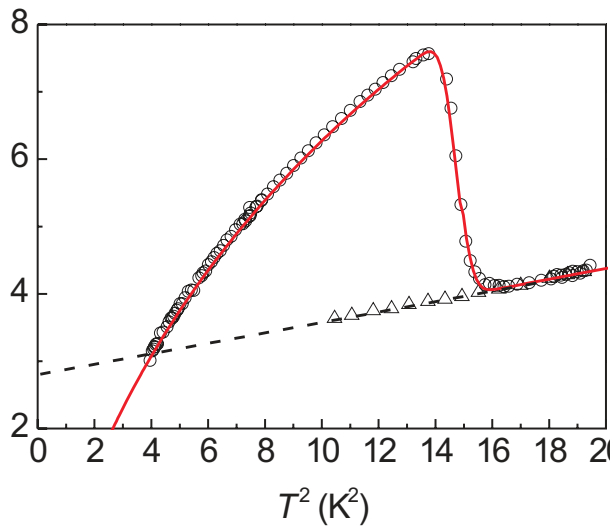}

\caption[]{Superconducting anomaly in the heat capacity of YC$_2$
($B_{\rm{ext}}$=0, $\circ$). The normal state
{{\scriptsize{$\bigtriangleup$}}} has been reached by applying an
exernal field of 0.4 T (after ref.[\onlinecite{GuldenPRB}]). The
(red) solid line represents a fit to the predictions of the BCS
theory with a slight smearing of $T_{\rm c}$ being included.}
\label{YC2CP}
\end{figure}

Heat capacity measurements (Fig.\ref{YC2CP}) nicely reveal the
anomaly at the transition to superconductivity which follows closely
the BCS weak-coupling predictions but already indicate significantly
decreased critical fields as compared to those of the layered
carbide halides. \cite{HennThesis}

\begin{figure}[h]
\includegraphics[width=8cm,angle=270]{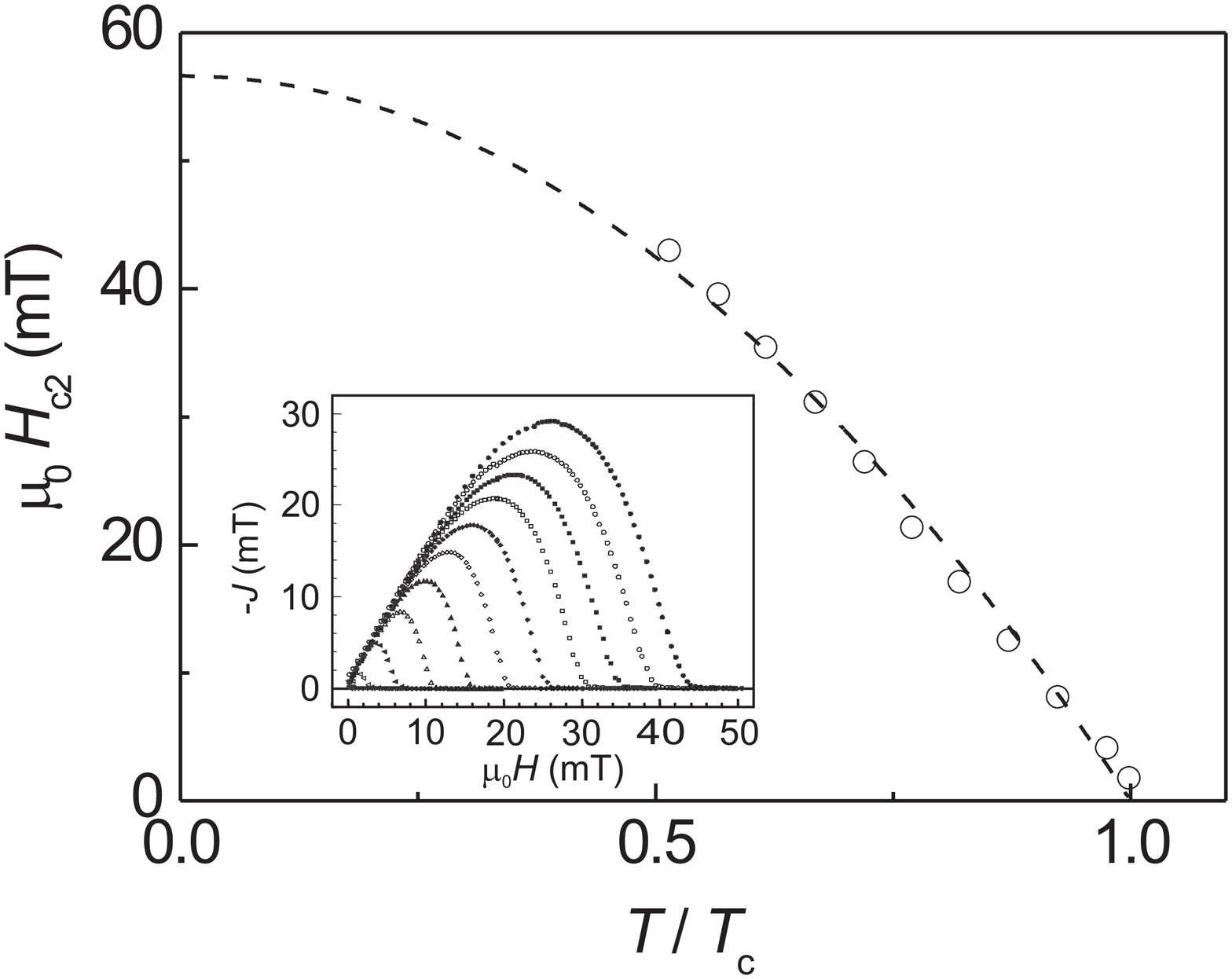}

\caption[]{Upper critical field, $\mu_0H_{\rm{c2}}$, determined from
the isothermal magnetization measured of a spherical sample of
YC$_2$. The inset displays the isothermal magnetizations measured at
constant temperatures of 2K, 2.2K, ..., 3.8K, 4K, in decreasing
order (after ref.[\onlinecite{HennThesis}]).} \label{YC2CRIT}
\end{figure}

Electronic structure calculations for YC$_2$ reveal strongly
dispersive bands in planes perpendicular to the {\it
c}\,-\,direction originating from Y $d_{x^2-y^2}$ orbitals and
also strongly dispersive bands in the {\it c}\,-\,direction
emerging from combinations of Y $d_{xz}$, $d_{yz}$, and C $p_x$,
$p_y$ orbitals. \cite{GuldenPRB} As a consequence the electronic
density of states close to the Fermi level is to a large extent
featureless with a slight positive slope. Doping with Th or Ca
(10\,\% to 20\,\%) decreases $T_{\rm c}$. \cite{GuldenPRB}

As compared to the layered yttrium carbide halides, the critical
fields of YC$_2$ ($<$\,0.1 T, cf. Fig.\ref{YC2CRIT} and Table
\ref{TableTc}) are reduced by up to two orders of magnitude . The
significant difference in the upper critical fields between the
layered carbide halides and the dicarbides as well as the marked
increase in the anisotropy of the coherence lengths
($\xi_\parallel / \xi_\perp \approx 5$, \cite{HennThesis})
supports Ginzburg's suggestion that from the point of view of
possibilities to enhance $T_{\rm c}$ promising materials are
layered materials and dielectric-metal-dielectric sandwich
structures. \cite{Ginzburg,Kirznits} In fact, by comparing the
crystal structures of the dicarbides and the carbide halides of
the rare earths (Figs. \ref{Y2STR} and \ref{YC2STR}) one realizes
that the R-C$_2$-R doublelayers carrying the superconductivity in
the ternary carbide halides can be considered as sections of the
three dimensional structure of the dicarbides which are sandwiched
by dielectric halogen layers. In this respect, the dicarbides and
the carbide halides of the rare earth metals are interesting
examples to test Ginzburg's conjecture.

\begin{figure}[h]
\includegraphics[width=10cm,angle=270]{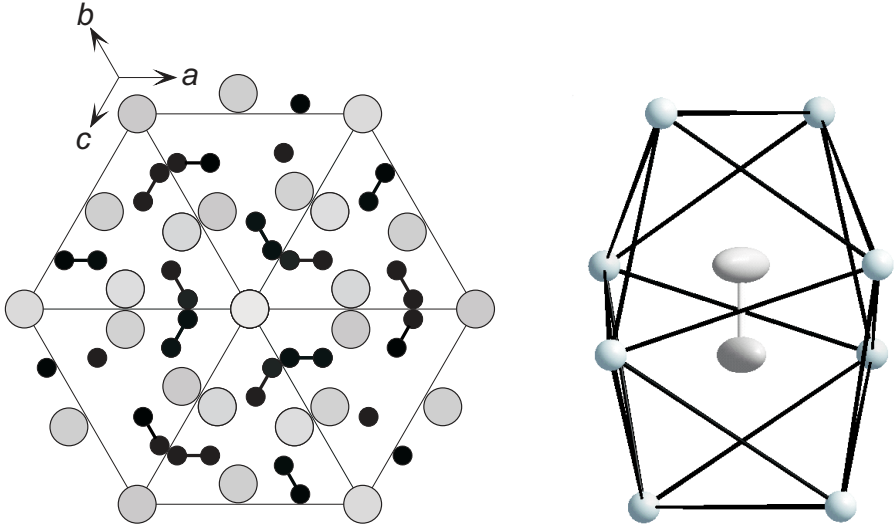}
\caption[]{Crystal structure of La$_2$C$_3$ projected along
[1\,1\,1] (after ref. [\onlinecite{GuldenThesis}]. (left) Unit
cell with La atoms indicated by the large spheres. (right) La atom
environment of a C\,-\,C dumbbell. The thermal ellipsoids of the C
atoms are shown.} \label{LA2STR}
\end{figure}

La$_2$C$_3$ (like Y$_2$C$_3$)  crystallizes with the cubic
Pu$_2$C$_3$ structure  in the space group $I$$\overline{\rm
4}$3$d$ which belongs to the tetrahedral crystallographic class
$T_d$ with no center of symmetry.\cite{Zachariasen} The structure
contains C--C dumbbells in a distorted dodecahedral coordination
(`bisphenoid') formed by 8 La atoms (cf. Fig. \ref{LA2STR}).  For
a more detailed discussion of the problems of C deficiency and the
problem of the aniostropy of the thermal ellipsoids of the C atoms
see ref. [\onlinecite{Kim2006}]. A recent study of the crystal
structure up to high pressures could not detect any structural
phase transitions up to 30 GPa. \cite{Wang}

In non-centrosymmetric systems with significant spin-orbit
coupling superconducting order parameters of different parity can
be mixed. A recent system which attracted particular interest in
this respect, is the heavy fermion superconductor CePt$_3$Si which
shows unconventional properties, as \textit{e.g.}
antiferromagnetism and superconductivity at $T_{\rm N}$$\sim$2.2 K
and $T_{\rm c}$ $\sim$ 0.75 K, respectively, and an upper critical
field  which considerably exceeds the paramagnetic limit.
\cite{Hilscher} With no 4$f$ electrons present and the high atomic
mass of La (as compared to Y) La$_2$C$_3$ is therefore an
interesting system to study the effects of non-centrosymmetry on
superconductivity. Possible multi-gap superconductivity is another
interesting issue which has been proposed for Th doped Y$_2$C$_3$
and La$_2$C$_3$. \cite{Sergienko} Recently, Harada \textit{et al.}
from $^{13}$C NMR measurements reported multi-gap
superconductivity for Y$_2$C$_3$.\cite{Harada}

\begin{figure}[h]
\includegraphics[width=12cm,angle=0]{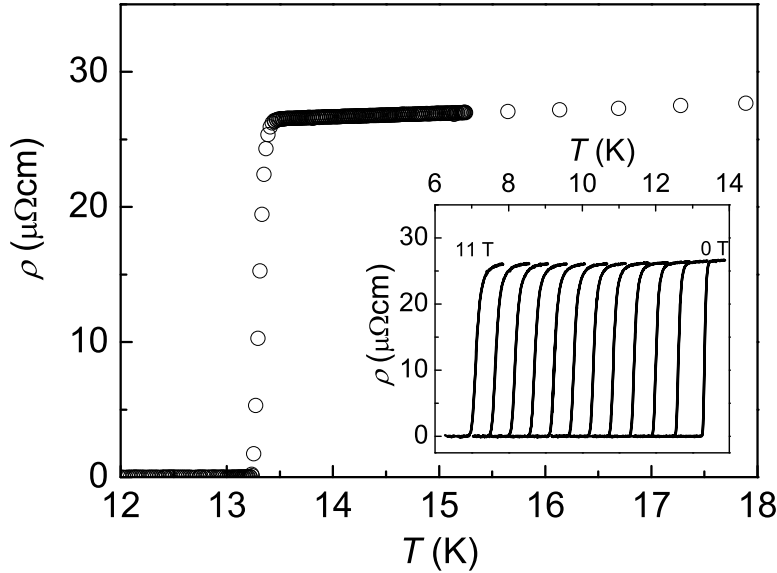}

\caption[]{Low temperature electrical resistivity of La$_2$C$_3$
showing the superconducting transition at 13.4 K. The inset
demonstrates the decrease of the superconducting transition with
external magnetic fields ranging from 0 T, 1 T, ..., 11 T.}
\label{LA2RHO}
\end{figure}

In contrast to Y$_2$C$_3$ which requires high-pressure synthesis
methods \cite{Amano,Nakane}, samples of La$_2$C$_3$ are readily
accessible by arc-melting of the constituents. Early on,
La$_2$C$_3$ was reported to have a $T_{\rm c}$ of $\sim$ 11 K.
\cite{Giorgi1969,Giorgi1970,Francavilla} Subsequently, it has been
shown that these samples were not stoichiometric, as anticipated,
but exhibit a range of homogeneity from 45.2\% to 60.2\% atom-\%
carbon content.\cite{GuldenThesis,Spedding,Simon2004}
Investigations of a series of samples La$_{2}$C$_{3-\delta}$ with
0.3 $\geq \delta \geq 0$ indicate a separation into two
superconducting phases with rather sharp $T_{\rm c}$'s of $\sim$ 6
K and 13.3\,-\,13.4 K (Fig. \ref{LA2RHO}). The high $T_{\rm c}$
values are attributed to stoichiometric La$_2$C$_3$, viz.
negligible C deficiency, which was assured individually for the
samples by neutron powder diffraction.
\cite{Kremer2004,Kim2006,Kim2007}

Our electronic structure calculations show a splitting of the
bands near $E_{\rm F}$ indicating that the spin degeneracy is
lifted due to a sizable spin-orbit coupling in addition to the
non-centrosymmetry in the structure.\cite{Kim2007} However, the
band splitting in La$_2$C$_3$ is much smaller than those found for
other non-centrosymmetric superconductors like CePt$_3$Si,
Li$_2$Pt$_3$B or Cd$_2$Re$_2$O$_7$. \cite{Samokhin,Lee,Eguchi} For
Li$_2$Pd$_3$B, another non-centrosymetric superconductor, where
the band splitting is comparable with that of La$_2$C$_3$,
conventional BCS type behavior with an isotropic superconducting
gap has been established via $\mu$SR experiments.\cite{Khasanov}
Based on our heat capacity measurements we similarly conclude that
La$_2$C$_3$ is a system with strong electron-phonon coupling with
a single gap of isotropic $s$-wave symmetry.

\begin{figure}[h]
\includegraphics[width=8cm,angle=0]{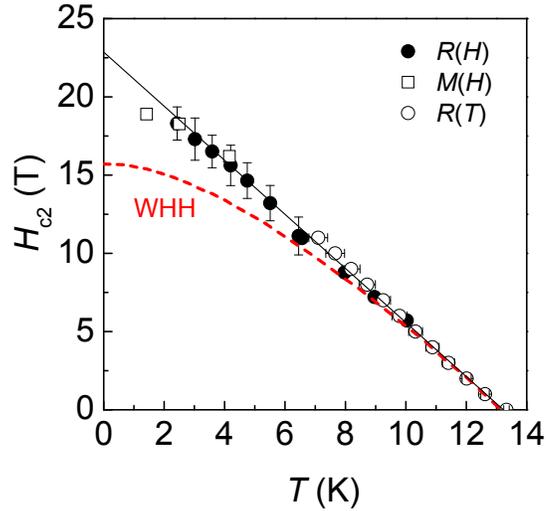}

\caption[]{Upper critical field of La$_2$C$_3$ determined by
various experimental methods, magnetoresistance ($R(H)$ and
$R(T)$) and magnetization ($M(H)$) measurements(after
ref.[\onlinecite{Kim2007}]).} \label{LA2CRIT}
\end{figure}

The upper critical field was determined by various methods and
reaches a value of $\sim$20~T at $T \rightarrow $ 0 K (Fig.
\ref{LA2CRIT}). $\mu_0 H_{\rm{c2}}$ is clearly enhanced over the
Werthamer-Helfand-\,Hohenberg predictions,\cite{WHH} but it does
not exceed the paramagnetic limit. Therefore, even though band
splitting effects due the non-centrosymmetric structure are
present, they appear to be not significant in case of La$_2$C$_3$.

\section{Superconductivity in Alkaline Earth Intercalated Graphite}

\begin{figure}[h]
\includegraphics[width=16cm,angle=0]{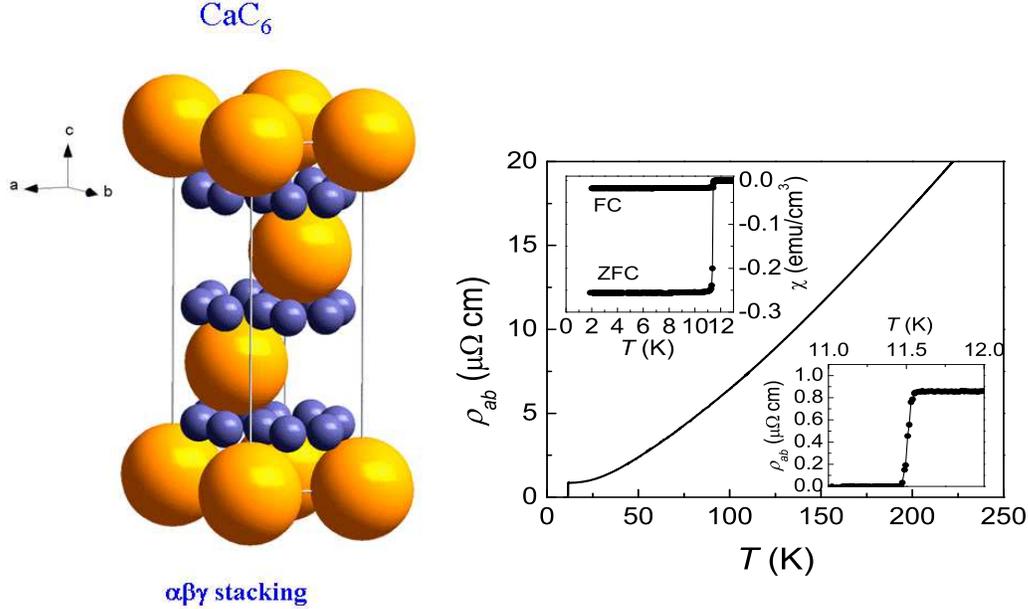}

\caption[]{(left) Crystal structure of CaC$_6$ (after
ref.[\onlinecite{Emery}]) and (right) in-plane electrical
resistivity and magnetic susceptibility (inset) of CaC$_6$ (after
ref.[\onlinecite{KIMPRL}]).} \label{CARHO}
\end{figure}

The recent discovery of superconductivity in Ca- and
Yb\,-\,intercalated graphite has refocused considerable interest
onto graphite intercalated compounds (GICs).\cite{Weller,Emery}
The superconducting transition temperatures for Ca- and Yb-
intercalated graphite are 11.5 and 6.5 K (cf. Fig. \ref{CARHO}),
respectively, significantly higher than that of the
alkali\,-\,metal intercalated graphite phases studied before.

Apart from the significant enhancement of $T_{\rm c}$, two other
aspects immediately attracted attention: In case of of YbC$_6$ it
was initially speculated that 4$f$ electrons may play a role and
that superconductivity might be mediated by valence fluctuation.
This possibility, however, could be ruled out and it was found
that Yb, like Ca, is divalent and the $f$ electrons provide no
essential contributions to the electronic structure at $E_{\rm
F}$. \cite{MazinMo} The second interesting aspect concerned the
role of the so-called `interlayer band', \textit{i.e.} a
three-dimensional nearly-free electron band emerging from
electrons localized in the intercalant plane, and its relation to
superconductivity together with the conjecture of an
unconventional electronic pairing mechanism involving
excitons.\cite{Csanyi}

This view was questioned based on the results of the first heat
capacity study on CaC$_6$. It  showed that the anomaly at $T_{\rm
c}$ can be clearly resolved indicating the bulk nature of the
superconductivity. In particular, both the temperature and
magnetic field dependence of $C_P$ strongly evidence a fully
gapped, intermediate-coupled, \textit{phonon-mediated}
superconductor without essential contributions from alternative
pairing mechanisms. \cite{KIMPRL}

\begin{figure}[h]
\includegraphics[width=9cm,angle=0]{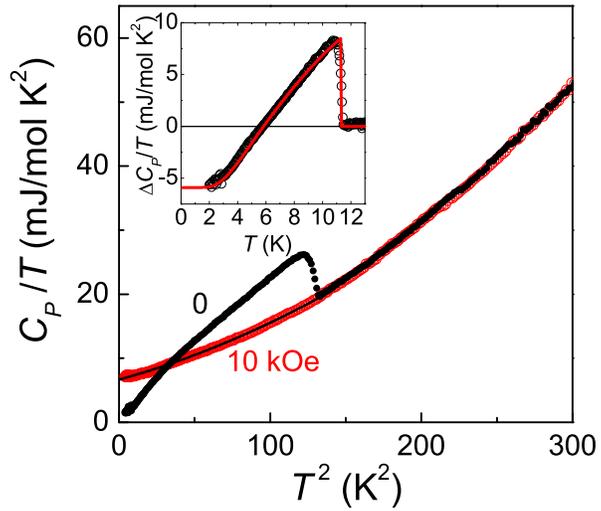}
\caption[]{Temperature dependence of the specific heat of CaC$_6$
at $B$ = 0 and 1 T. the inset shows the temperature dependence of
$\Delta C_P / T$ = $C_P / T (B = 0)$ - $C_P / T (B = 1{\rm{T}})$.
The (red) solid line is the best fit assuming  an isotropic
$s$-wave BCS gap. (after ref.[\onlinecite{KIMPRL}]).} \label{CACP}
\end{figure}

\begin{figure}[h]
\includegraphics[width=10cm,angle=0]{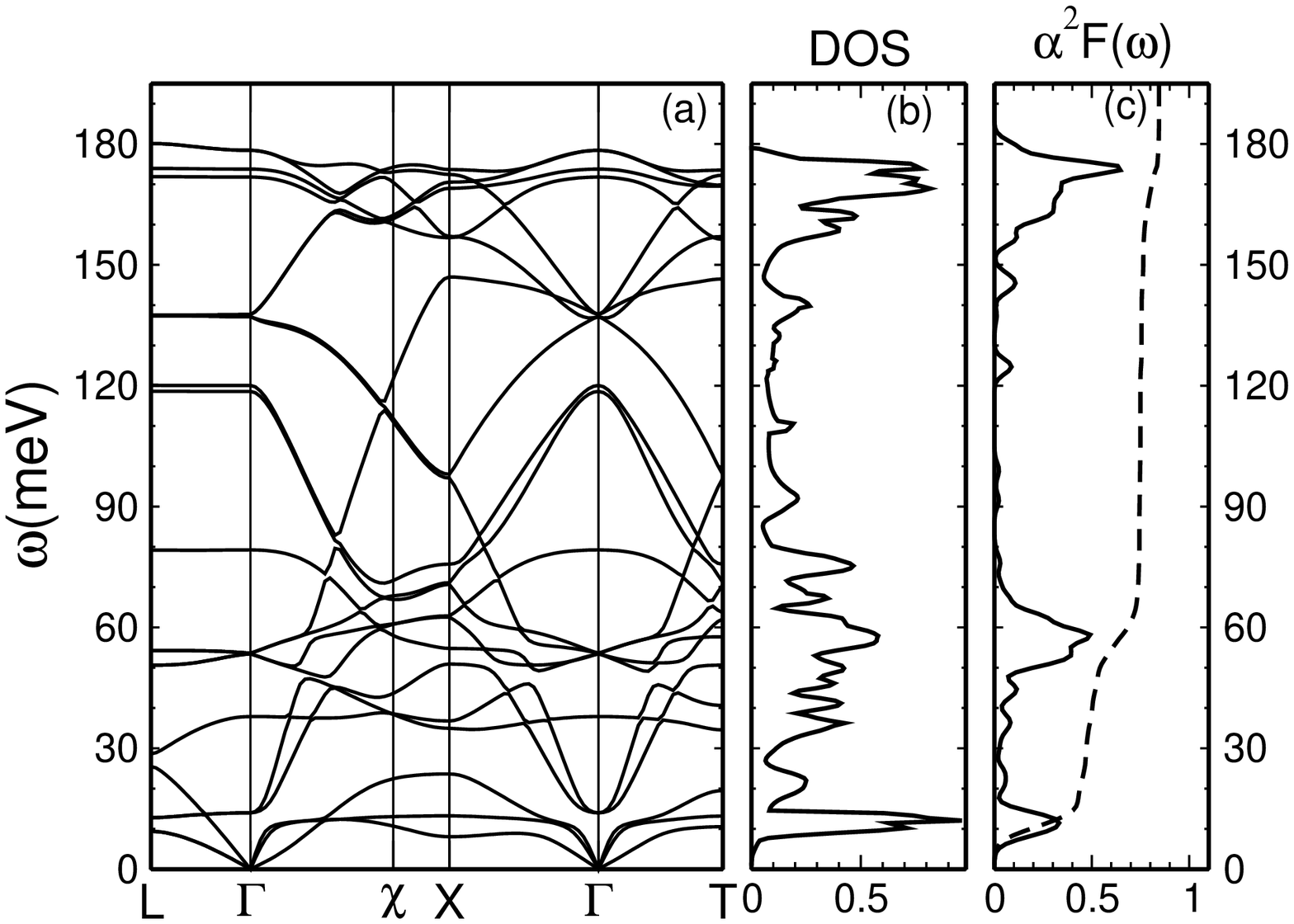}
\caption[]{(a) Phonon frequencies and (b) density of states of
CaC$_6$ along selected directions in the rhombohedral unit cell;
the line $\Gamma$ - $X$ is contained in the graphene planes, while
$L$ - $\Gamma$ is orthogonal to it. (c) Eliashberg function
$\alpha^2 F(\omega)$ and frequency - dependent electron - phonon
coupling $\lambda(\omega)$ (after ref.[\onlinecite{KimPRB2007}]).}
\label{CAPHO}
\end{figure}

Linear response calculations provide the following picture of the
electron\,-\,phonon coupling (cf. Fig. \ref{CAPHO}): There are
three distinct groups of modes, one at $\omega \sim$ 10 meV,
another around $\omega \sim$ 60 meV, and the third located at
$\omega \sim$ 170 meV which contribute $\sim$0.4, $\sim$0.3, and
$\sim$0.1 to the total coupling constant $\lambda$. These three
groups are mainly composed of the Ca, out-of plane and in-plane C
vibrations, respectively. The observed positive pressure
dependence of $T_{\rm c}$ can be understood within this
electron\,-\,phonon coupling scheme due to a softening of the Ca
in-plane phonon modes.\cite{Mauri,KimPRB2007}

\section{Conclusions}
 The broad chemical bonding abilities
of carbon allowing to realize highly anisotropic chemical
structures which together with the low atomic mass of carbon make
the modifications of carbon as well as carbon\,-\,derived
compounds to a wide and rewarding playground to search for new and
unusual superconductors.

\end{document}